\documentclass[pdflatex,sn-aps, iicol]{sn-jnl}% American Physical Society (APS) Reference Style
\usepackage{graphicx}%
\usepackage{multirow}%
\usepackage{amsmath,amssymb,amsfonts}%
\usepackage{amsthm}%
\usepackage{mathrsfs}%
\usepackage[title]{appendix}%
\usepackage{xcolor}%
\usepackage{textcomp}%
\usepackage{manyfoot}%
\usepackage{booktabs}%
\usepackage{algorithm}%
\usepackage{algorithmicx}%
\usepackage{algpseudocode}%
\usepackage{listings}%
\usepackage{graphicx}
\usepackage{subcaption}
\theoremstyle{thmstyleone}%
\theoremstyle{thmstyletwo}%

\theoremstyle{thmstylethree}%

\begin{document}

\title[Cosmic anisotropy with GRBs]{Testing cosmic anisotropy with the Combo correlation of gamma-ray bursts}

\author[1]{\fnm{Dong} \sur{Zhao}}%\email{iauthor@gmail.com}

\author[1]{\fnm{Hao-Ran} \sur{Duan}}%\email{iiauthor@gmail.com}

\author*[2,3]{\fnm{Jun-Qing} \sur{Xia}}\email{xiajq@bnu.edu.cn}

\affil[1]{Center for Gravitation and Cosmology, College of Physical Science and Technology, Yangzhou University, Yangzhou, 225009, China}

\affil[2]{Institute for Frontiers in Astronomy and Astrophysics, Beijing Normal University, Beijing, 100875, China}

\affil[3]{School of Physics and Astronomy, Beijing Normal University, Beijing, 100875, China}

%Center for Gravitation and Cosmology, College of Physical Science and Technology, Yangzhou University, Yangzhou, 225009, China

\abstract{
	
We employ the sample of 244 gamma-ray bursts (GRBs; i.e., C244) with the Combo correlation to test cosmic anisotropy. Meanwhile, the Pantheon sample is introduced to verify whether the GRB sample can suppress the fake anisotropic signals induced by inhomogeneous spatial distributions. In the dipole fitting (DF) method, under the dipole-modulated $\Lambda$CDM model, the C244 sample shifts the best-fitting longitude $l$ derived from the Pantheon sample by $54.09^\circ$ and reduces the uncertainty in $l$ by approximately $40\%$. Compared to the 118 GRBs (i.e., A118) with the $E_\mathrm{p}$-$E_\mathrm{iso}$ correlation, the shift in longitude $l$ increases by additional $21.35^\circ$. In the hemisphere comparison (HC) method, the preferred direction derived from the C244+Pantheon sample deviates from that of the Pantheon-only sample by more than $1\sigma$. In contrast, the preferred direction from the A118+Pantheon sample is consistent with the Pantheon-only result within the $1\sigma$ uncertainty. The preferred direction changes significantly as the number of GRBs increases from 118 to 244. Our results show that a larger GRB sample can reduce the fake anisotropic signals caused by inhomogeneous spatial distributions. Accordingly, we suggest that GRBs have the potential to provide a reliable probe of cosmic anisotropy.

}

%\keywords{keyword1, Keyword2, Keyword3, Keyword4}

\maketitle

\section{Introduction}\label{sec1}

The theoretical framework of modern cosmology is based on the cosmological principle, which assumes that the Universe exhibits homogeneity and isotropy at large scales. Observations of the cosmic microwave background (CMB) radiation by the Wilkinson Microwave Anisotropy Probe (WMAP) \citep{WMAP:2012nax} and Planck satellites \citep{Planck:2018vyg} have confirmed its validity with high precision. However, several anomalies emerge from current observations, posing a potential challenge to the cosmological principle. These include the hemispherical power asymmetry \citep{Planck:2019evm} and the parity asymmetry \citep{Kim:2010gf, Gruppuso:2010nd, Zhao:2013jya} in the CMB, as well as the large-scale alignment of quasar polarization vectors \citep{Hutsemekers:2005iz, Pelgrims:2016zbr}. These anomalies suggest that the Universe may have a preferred direction.

As standard candles in cosmology, datasets of Type Ia supernovae (SNe Ia) have been extensively used to test the cosmological principle. These datasets include Union2 \citep{Amanullah:2010vv} and 2.1 \citep{SupernovaCosmologyProject:2011ycw}, JLA \citep{Betoule:2014frx}, Pantheon \citep{Scolnic:2017caz}, and Pantheon+ \citep{Scolnic:2021amr}. Studies have shown that although a dipole signal with 2$\sigma$ statistical significance was detected in the Union2 \citep{Mariano:2012wx, Zhao:2013yaa} and 2.1 \citep{Yang:2013gea, Li:2015uda, Lin:2016jqp, Sun:2018epo} datasets, this dipole signal has not been confirmed in subsequent JLA \citep{Lin:2015rza, Wang:2017ezt, Chang:2017bbi}, Pantheon \citep{Sun:2018cha, Deng:2018jrp, Zhao:2019azy, Chang:2019utc}, and Pantheon+ \citep{Tang:2023kzs} datasets. However, some studies have pointed out that a significant dipole signal can still be detected in the low-redshift subsample of the Pantheon+ sample \citep{Sorrenti:2022zat, Cowell:2022ehf, Tang:2023kzs, Sorrenti:2024ugq, Sah:2024csa}. Furthermore, when analyzed using the HC method, both the Pantheon \citep{Zhao:2019azy} and Pantheon+ \citep{Hu:2023eyf, Hu:2024qnx} datasets exhibited significant anisotropic signals. There is a sharp change in the anisotropy level of the Pantheon+ sample occurring at distances less than 40$Mpc$ \cite{Perivolaropoulos:2023tdt}. It is also worth noting that some studies suggest that the Pantheon+ sample are consistent with the isotropic assumption \citep{Bengaly:2024ree, Zhou:2025rvf}.

In addition to SNe Ia datasets, other types of cosmological observations are also widely used to test cosmic anisotropy. An incomplete list of related works includes galaxies \citep{Boubel:2024cmh, Stiskalek:2025cjv}, galaxy clusters \citep{Migkas:2017vir, Migkas:2020fza, Migkas:2021zdo, Aluri:2022hzs, Pandya:2024jqg}, gravitational waves \citep{Cai:2019cfw, Cousins:2024bhk, Chen:2025qsl, Wang:2025ygl}, fast radio bursts \citep{Cai:2019cfw, Qiang:2019zrs, Lin:2021syj}, the sky distribution of radio and infrared sources \citep{Secrest:2020has, Secrest:2022uvx, Dam:2022wwh, Guandalin:2022tyl, Singal:2023wni, Wagenveld:2023kvi, Cheng:2023eba, Panwar:2023uqf, Singal:2023lqm, daSilveiraFerreira:2024ddn, Singal:2024ldf, Abghari:2024eja, Tiwari:2024dsj, Wagenveld:2025ewl}, quasars \citep{Hu:2020mzd, Zhao:2021zih, Zhao:2021fcp}, and GRBs \cite{Li:2015yha, Chang:2014jza, Wang:2014vqa, Luongo:2021nqh, Zhao:2022dwy, Luongo:2025kfs, Santiago:2025rmf}.

Some studies suggest that anisotropic signals may stem from the inhomogeneous spatial distribution of datasets in the sky. Zhao et al. \citep{Zhao:2019azy} discovered that the anisotropic signals detected in the Pantheon sample are heavily dependent on the inhomogeneous distribution of SNe Ia in the sky. Sun and Wang \cite{Sun:2018epo} found that the anisotropic distribution of coordinates can cause the dipole direction to change and increase the magnitude of the dipole. Therefore, the search for convincing anisotropic signals requires observational datasets with a more uniform spatial distribution.

The spatial distribution of quasar samples \cite{Risaliti:2015zla, Risaliti:2018reu, Lusso:2020pdb} is more uniform than that of the Pantheon sample. Some studies have used quasar samples to investigate anisotropic signals \citep{Hu:2020mzd, Zhao:2021zih, Zhao:2021fcp}. However, it should be noted that the “more uniform” spatial distribution of quasar samples is only relative to the Pantheon sample. The quasar samples still exhibit significant spatial inhomogeneities. For the 2020 compilation of quasars \citep{Lusso:2020pdb}, almost 70\% of quasars are located in the northern galactic hemisphere, and the rest are concentrated in the southeastern galactic hemisphere, meaning that their spatial distribution remains highly inhomogeneous. In contrast, GRBs are the most powerful explosions observed in the universe, with a redshift range extending up to z = 9.4 \citep{Cucchiara:2011pj, DAvanzo:2012qsl}. The peak of their redshift distribution is almost at the maximum redshift of SNe Ia \citep{Coward:2012ay}. More importantly, the spatial distribution of GRBs is nearly homogeneous.

Previously, we tested cosmic anisotropy with the $E_\mathrm{p}$-$E_\mathrm{iso}$ (`Amati') correlation \citep{Amati:2018tso} of 118 long GRBs \citep{Dirirsa:2019fcs, Khadka:2021vqa}. We found that although the number of GRBs is only about a tenth of the Pantheon sample size, they can considerably impact the anisotropic signals arising from the inhomogeneous sky distribution of the Pantheon sample \citep{Zhao:2022dwy}. In this work, we will analyze a larger sample of 244 GRBs \citep{Muccino:2026gvt} with the Combo correlation \citep{Izzo:2015vya, Muccino:2020gqt}. We will perform a joint analysis by combining the Pantheon sample with these 244 GRBs. Because the anisotropic signals in the Pantheon sample stem from the uneven spatial distribution of its data across the sky \citep{Zhao:2019azy}, it serves as an ideal tool to test whether introducing a more homogeneously distributed dataset can suppress these signals. We will investigate the impact of expanding the GRB sample size from 118 to 244 on the anisotropic signals in the Pantheon sample. The rest of this paper is organized as follows. In Section \ref{Methodology}, we briefly introduce the 244 GRBs and the Combo correlation. In Section \ref{Results}, we employ the DF and HC methods to investigate cosmic anisotropy. Finally, conclusions and discussions are presented in Section \ref{Conclusion}.

\section{Methodology}\label{Methodology}
\subsection{The GRB sample}\label{subsec1}
\begin{figure}
	\begin{center}
		\includegraphics[width=7.5cm]{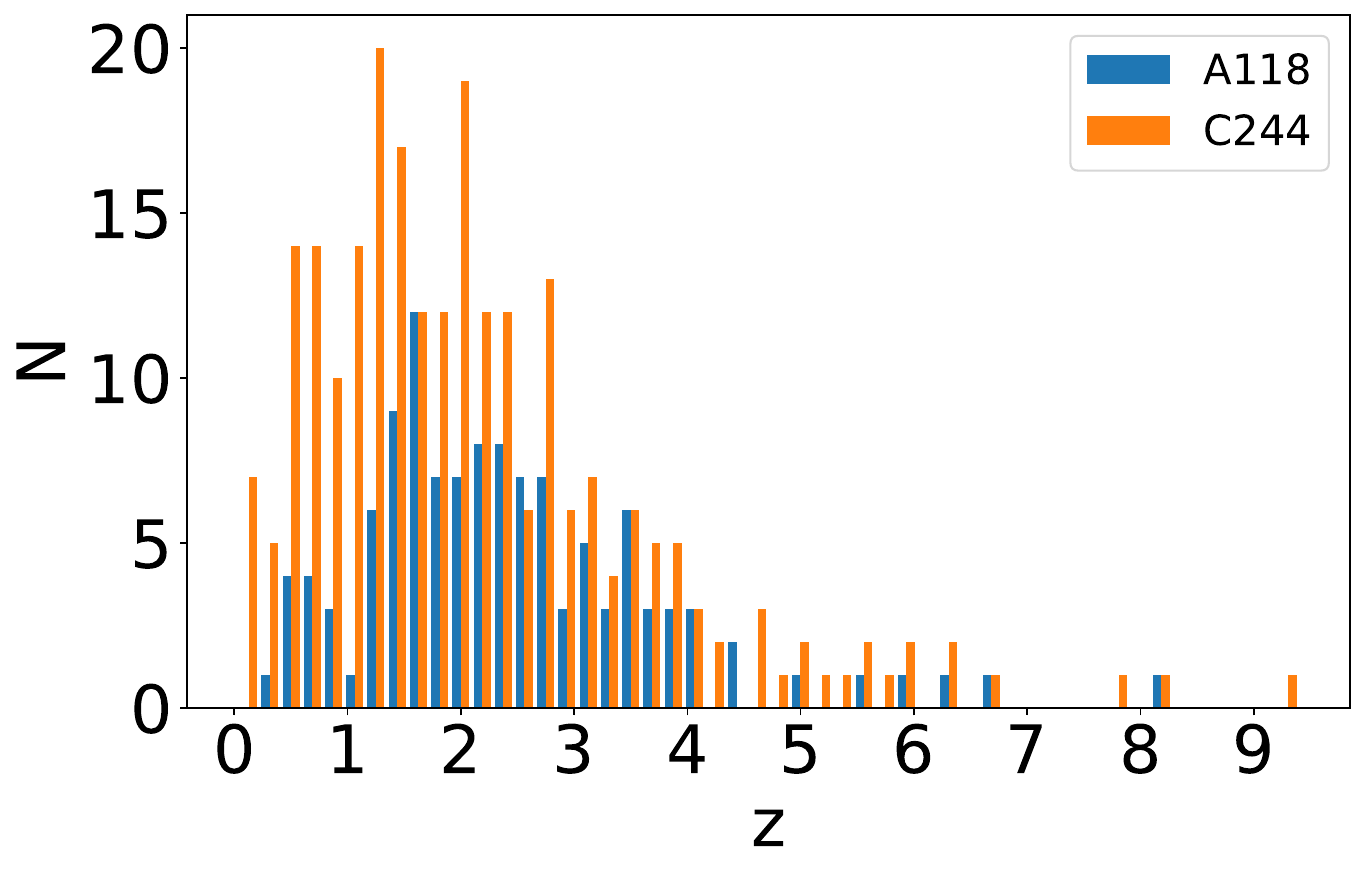}
		\caption{Redshift distribution of the GRB samples. The blue histogram denotes the A118 sample \citep{Dirirsa:2019fcs, Khadka:2021vqa}, while the orange histogram represents the C244 sample \citep{Muccino:2026gvt}.}
		\label{fig:red_GRB}
	\end{center}
\end{figure}

\begin{figure}
	\begin{center}
		\includegraphics[width=7.5cm]{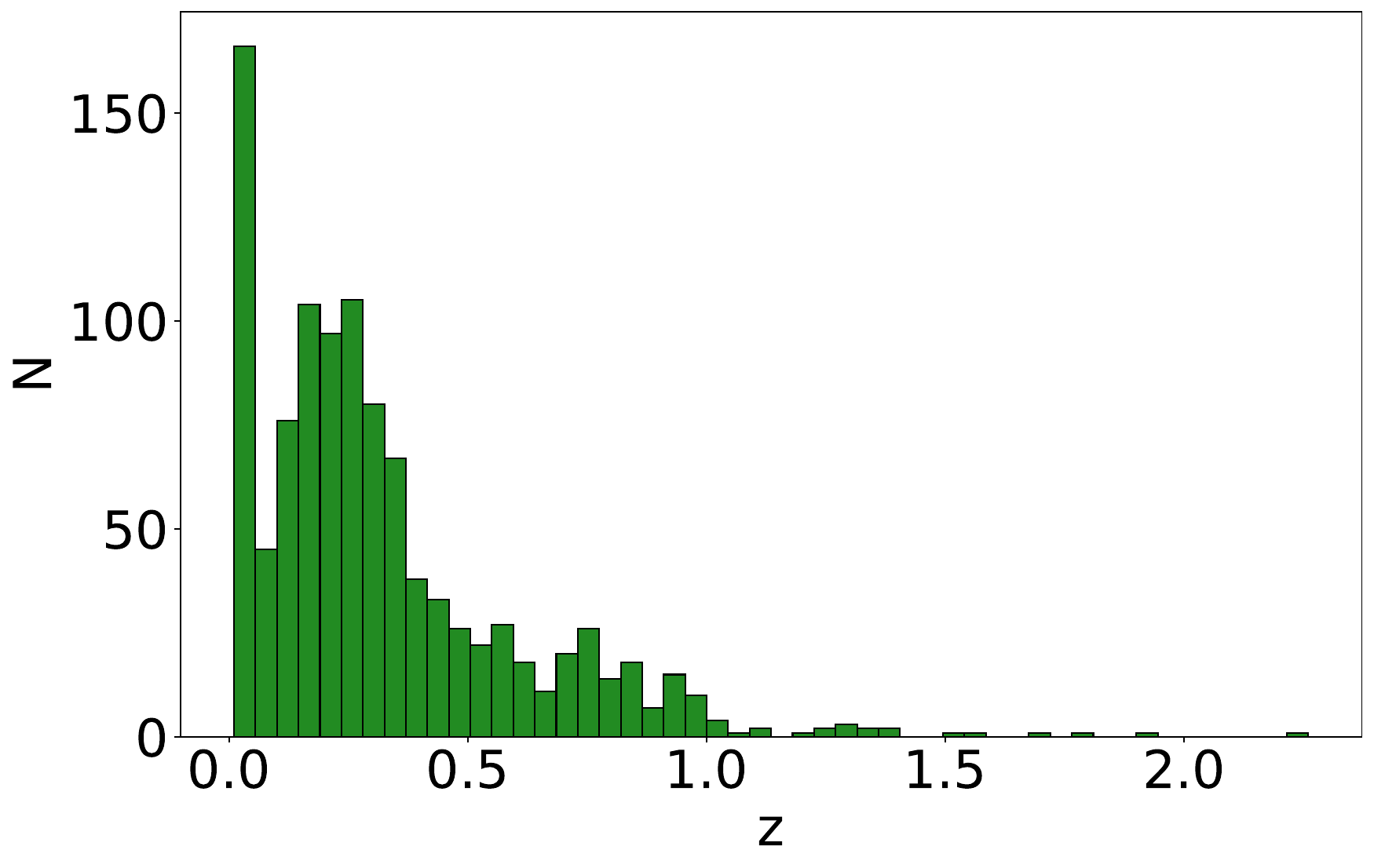}
		\caption{Redshift distribution of the Pantheon sample.}
		\label{fig:red_P}
	\end{center}
\end{figure}

\begin{figure*}[htbp]
	\centering
	\begin{subfigure}[b]{15cm}
		\centering
		\includegraphics[width=\textwidth]{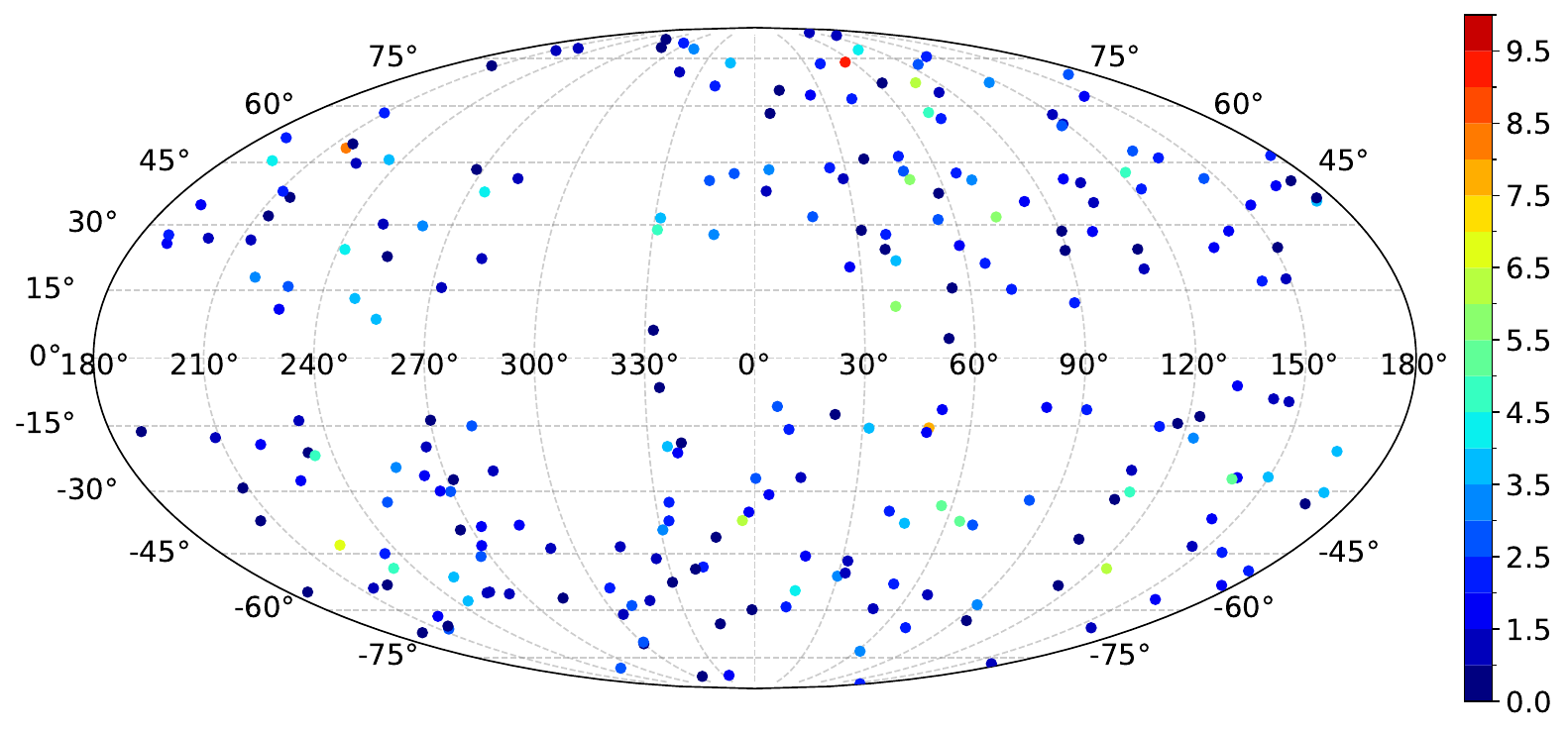}
		\caption{C244 sample}
		\label{fig:lb_GRB_C}
	\end{subfigure}
	\\ \vspace{10pt}
	\begin{subfigure}[b]{15cm}
		\centering
		\includegraphics[width=\textwidth]{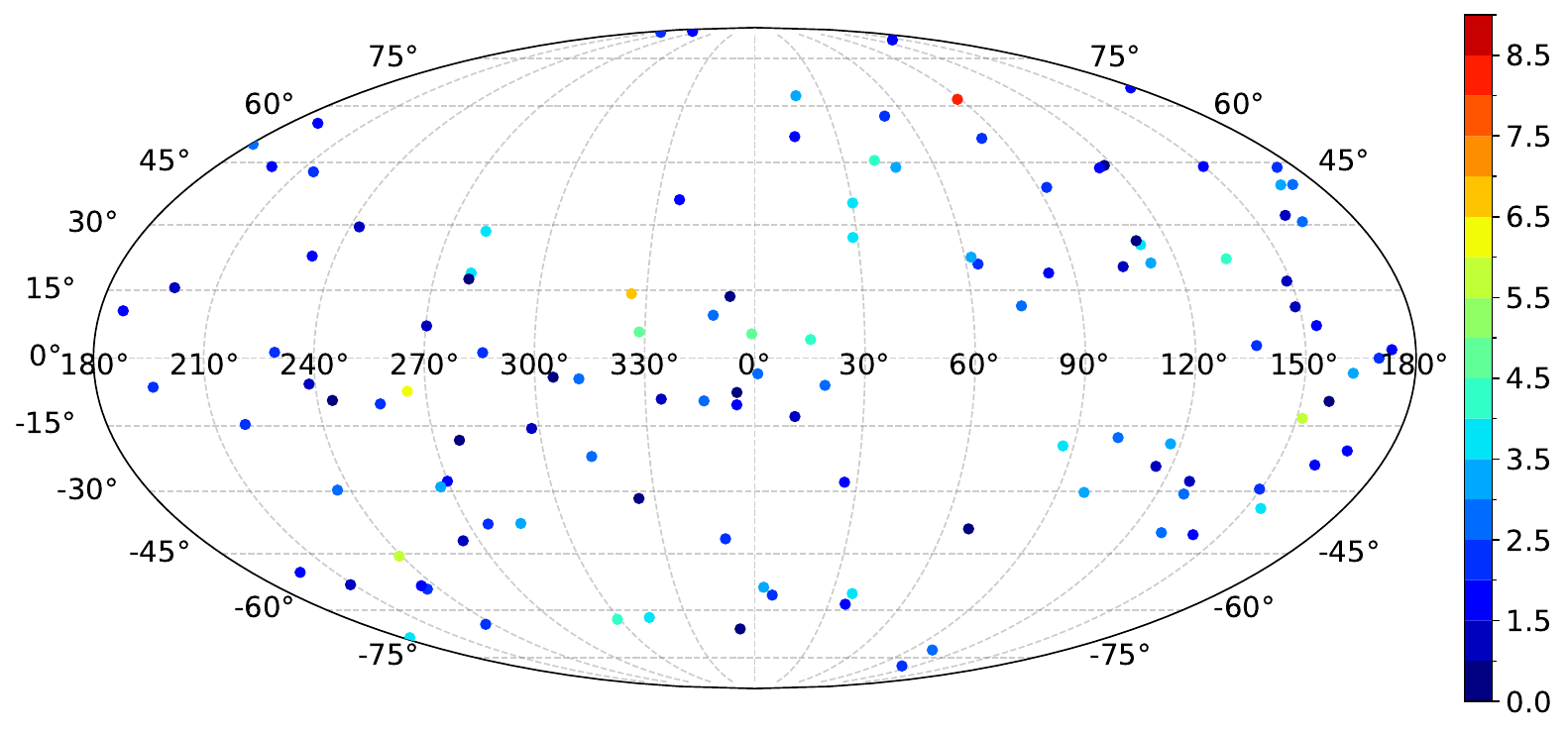}
		\caption{A118 sample}
		\label{fig:lb_GRB_A}
	\end{subfigure}
	\\ \vspace{10pt}
	\begin{subfigure}[b]{15cm}
		\centering
		\includegraphics[width=\textwidth]{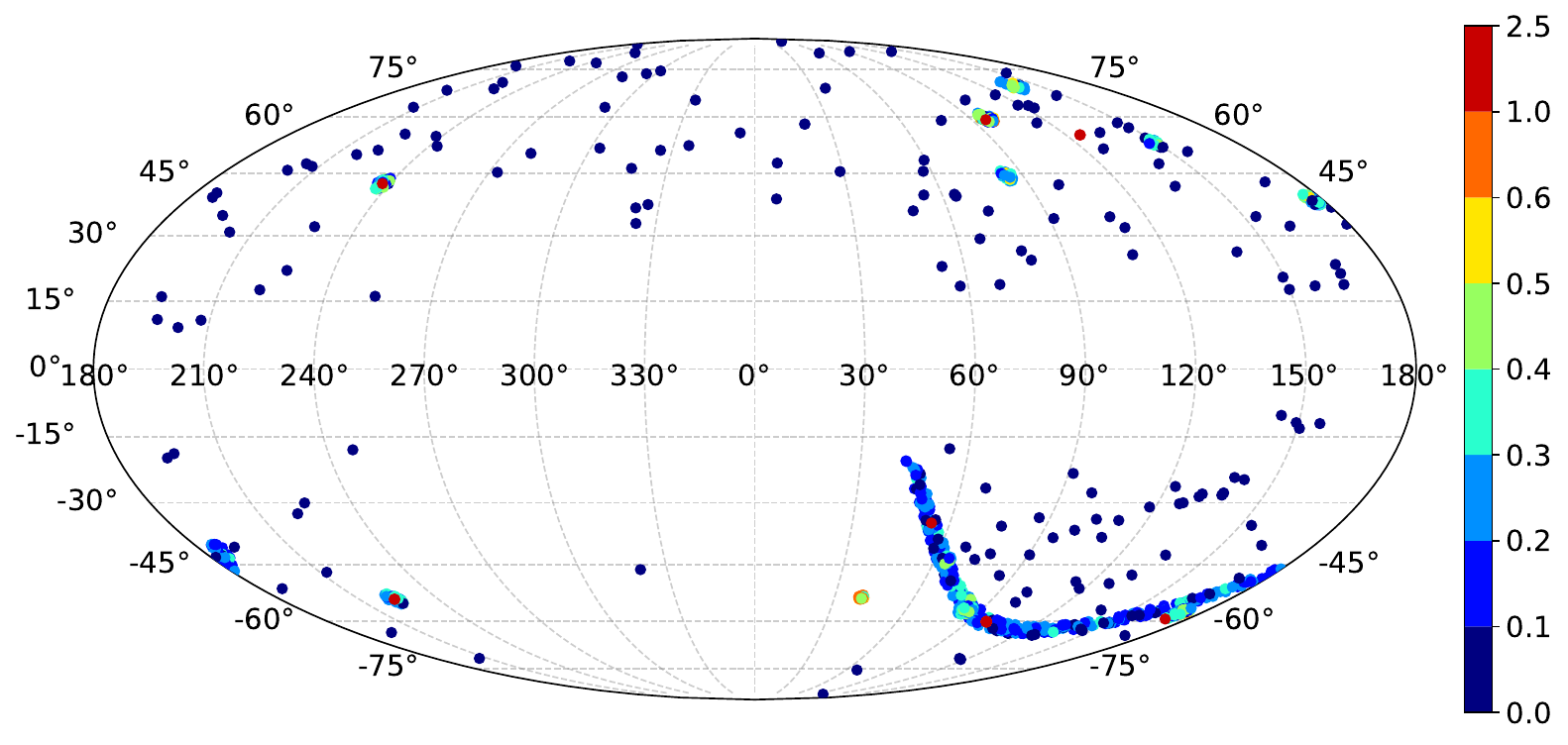}
		\caption{Pantheon sample}
		\label{fig:lb_P}
	\end{subfigure}
	\caption{Distributions of the GRB samples and the Pantheon sample in Galactic coordinates, where the pseudo-colors represent the redshift of the sources.}
	\label{fig:lb_distribution} 
\end{figure*}

The 244 GRBs \citep{Muccino:2026gvt} used in our work span a redshift range of $ 0.0331 \leq z \leq 9.4$. Fig. \ref{fig:red_GRB} displays their redshift distribution (orange histogram). For comparison, we also show the redshift distribution of the A118 sample used in our previous work \citep{Zhao:2022dwy} in Fig. \ref{fig:red_GRB} with the blue histogram. Most C244 sources are concentrated within $ 0 < z < 4$. The C244 sample exhibits a higher density of GRBs in the redshift range $ 0 < z < 3$, and extends up to $z = 9.4$. The redshift distribution of the Pantheon sample is shown in Fig. \ref{fig:red_P}. We show the spatial distributions of the C244, A118, and Pantheon samples in Fig. \ref{fig:lb_distribution}. The  C244 and A118 samples are both nearly uniformly distributed in the sky. In contrast, the Pantheon sample is highly inhomogeneous, with approximately 335 sources concentrated in a distinct band along the celestial equator.

\subsection{The Combo correlation}\label{subsec2}
The Combo correlation \citep{Izzo:2015vya, Muccino:2020gqt} is an empirical relationship that combines prompt and afterglow emissions in long GRBs, expressed as
\begin{equation}\label{equ_Combo}
	\log\left(\frac{L_0}{\mathrm {erg/s}}\right) = \alpha + \beta \log\left(\frac{E_\mathrm {p}}{\mathrm {keV}}\right) - \log\left(\frac{T}{s}\right)\ ,
\end{equation}
where $\log$ denotes $\log_{10}$. $\alpha$ and $\beta$ are two free parameters determined by the C244 sample. $E_\mathrm{p}$ is the rest-frame peak energy of the GRB photon energy spectrum. $T$ is the X-ray afterglow rest-frame effective duration of the plateau phase, and $L_0$ represents the X-ray afterglow plateau luminosity, which can be derived from
\begin{equation}\label{th_L0}
	L_0 = 4\pi d_\mathrm{L}^2(z) F_0,
\end{equation}
where $F_0$ denotes the X-ray afterglow flux in the rest frame. The luminosity distance $d_\mathrm{L}$ has the form
\begin{equation}\label{dL}
	d_\mathrm{L}=\frac{c(1+z)}{H_0} \int_{0}^{z} \frac{d z^{\prime}}{E\left(z^{\prime}\right)},
\end{equation}
$z$ represents the redshift of sources, and $c$ is the speed of light. $H_{0}$ denotes the Hubble constant. In this work, $H_{0}$ is fixed to $70 \mathrm{~km} \mathrm{~s}^{-1} \mathrm{Mpc}^{-1}$. In the flat $\Lambda$CDM model, $E(z)$ is given as
\begin{equation}
	E(z)=\sqrt{\Omega_\mathrm{m}(1+z)^{3}+\left(1-\Omega_\mathrm{m}\right)},
\end{equation}
where $\Omega_\mathrm{m}$ is the matter density.

The $\chi^2$ of the GRB sample can be formulated as \citep{DAgostini:2005mth}
\begin{equation}\label{chi_GRB}
	\chi^2_\mathrm{GRB}=\sum_{i=1}^{N_\mathrm{GRB}}\left[\frac{\left(\log L_{0, i}^{\mathrm{th}}-\log L_{0, i}^{\mathrm{obs}}\right)^{2}}{s_{i}^{2}}+\ln \left(2 \pi s_{i}^{2}\right)\right],
\end{equation}
where $s_{i}^{2}$ is computed via
\begin{equation}
	s_{i}^2=\sigma_{\log{F_{0,i}}}^2+\beta^2\sigma_{\log{E_{\mathrm{p},i}}}^2+\sigma_{\log{T_{i}}}^2+\sigma^2,
\end{equation}
where $\sigma_{\log{F_{0}}}$, $\sigma_{\log{E_\mathrm{p}}}$, and $\sigma_{\log{T}}$ represent the uncertainties of $\log{F_{0}}$, $\log{E_\mathrm{p}}$, and $\log{T}$, respectively. $\sigma$ denotes systematics and additional hidden uncertainties.

\subsection{The dipole fitting method}\label{subsec3}
Mariano and Perivolaropoulos \cite{Mariano:2012wx} proposed the DF method. It is now widely used to test cosmic anisotropy. In this work, we employ a distance modulus modified by a dipole modulation, expressed as
\begin{equation}\label{dm_dipole}
	\tilde{\mu}_\mathrm{th}=\mu_\mathrm{th} \times\left(1+A_\mathrm{D}(\hat{\boldsymbol{n}} \cdot \hat{\boldsymbol{p}})\right),
\end{equation}
where $\hat{\boldsymbol{n}}$ is the dipole direction and $\hat{\boldsymbol{p}}$ is the unit vector pointing to the position of the sources. $A_\mathrm{D}$ denotes the dipole amplitude. In Galactic coordinates, the dipole direction $\hat{\boldsymbol{n}}$ is parameterized as
\begin{equation}
	\hat{\boldsymbol{n}}=\cos (b) \cos (l) \hat{\boldsymbol{i}}+\cos (b) \sin (l) \hat{\boldsymbol{j}}+\sin (b) \hat{\boldsymbol{k}},
\end{equation}
where $l$ and $b$ denote the Galactic longitude and latitude of the dipole direction, respectively. $\hat{i}$, $\hat{j}$, and $\hat{k}$ are the standard Cartesian unit vectors. The unit vector $\hat{\boldsymbol{p}}$ points to the $i$-th source, taking the form
\begin{equation}
	\hat{\boldsymbol{p}}_{i}=\cos \left(b_{i}\right) \cos \left(l_{i}\right) \hat{\boldsymbol{i}}+\cos \left(b_{i}\right) \sin \left(l_{i}\right) \hat{\boldsymbol{j}}+\sin \left(b_{i}\right) \hat{\boldsymbol{k}}.
\end{equation}

In the $\Lambda$CDM model, the theoretical distance modulus is given by
\begin{equation}\label{dis_modu}
\mu_\mathrm{th}=5 \log \frac{d_\mathrm{L}}{\mathrm{Mpc}}+25,
\end{equation}
By combining Eqs. (\ref{dm_dipole}) and (\ref{dis_modu}), we derive the expression for the luminosity distance under dipole modulation, expressed as
\begin{equation}\label{t_log}
	\log \frac{\tilde{{d_\mathrm{L}}}}{\mathrm{Mpc}}=\left(\log \frac{d_\mathrm{L}}{\mathrm{Mpc}}+5\right) \times(1+A_\mathrm{D}(\hat{\mathbf{n}} \cdot \hat{\mathbf{p}}))-5,
\end{equation}
plugging Eq. (\ref{t_log}) into Eq. (\ref{th_L0}), we obtain the expression for the X-ray afterglow plateau luminosity $L_0$ under dipole modulation.

In addition to the dipole-modulated $\Lambda$CDM model, we also explore the Finslerian cosmological model \citep{Li:2015uda} in our analysis. The dipole structure is naturally given in the expression of $E(z)$, which takes the form
\begin{equation}
	E(z)=\sqrt{\Omega_\mathrm{m}(1+z)^{3}(1 + A_\mathrm{D} \cos \theta)^{-3}+1-\Omega_\mathrm{m}},
\end{equation}
where $A_{D}$ denotes the dipole amplitude and $\theta$ is the angle between the dipole direction and the position of the sources.

\subsection{The hemisphere comparison method}\label{subsec4}
Schwarz and Weinhorst \cite{Schwarz:2007wf} originally introduced the HC method to probe cosmic anisotropy. The steps of the HC method can be summarized as:

~~(1) Generate a random direction $\hat{\boldsymbol{n}}(l, b)$ in Galactic coordinates, where $l$ and $b$ represent the Galactic longitude and latitude, respectively. Given this random direction $\hat{\boldsymbol{n}}$, the celestial sphere is divided into two distinct ``up'' and ``down'' hemispheres.

~~(2) Based on the sources' coordinates, the sample can be divided into two subsets corresponding to the ``up'' and ``down'' hemispheres, respectively. 

~~(3) Find the best-fitting value of $\Omega_\mathrm{m}$ and its $1\sigma$ uncertainty for each subset assuming a flat $\Lambda$CDM model. The anisotropy level (AL) is defined as
\begin{equation}
	\mathrm{AL}=2 \times \frac{\Omega_\mathrm{m, u}-\Omega_\mathrm{m, d}}{\Omega_\mathrm{m, u}+\Omega_\mathrm{m, d}},
\end{equation}
where the subscript $u$ denotes the ``up'' hemisphere and the subscript $d$ denotes the ``down'' hemisphere. The $1\sigma$ uncertainty of AL is given by
\begin{equation}
	\sigma_{\mathrm{AL}}=\frac{\sqrt{\sigma_{\Omega_\mathrm{m, u}}^{2}+\sigma_{\Omega_\mathrm{m, d}}^{2}}}{\Omega_\mathrm{m, u}+\Omega_\mathrm{m, d}}.
\end{equation}
~~(4) Repeat the above steps to generate adequate directions. Find the maximum AL (i.e., $\mathrm{AL}_\mathrm{max}$) and its corresponding direction.

\section{Results}\label{Results}

\subsection{Dipole fitting}\label{R_DF}
\begin{figure*}
	\begin{center}
		\includegraphics[width=15cm]{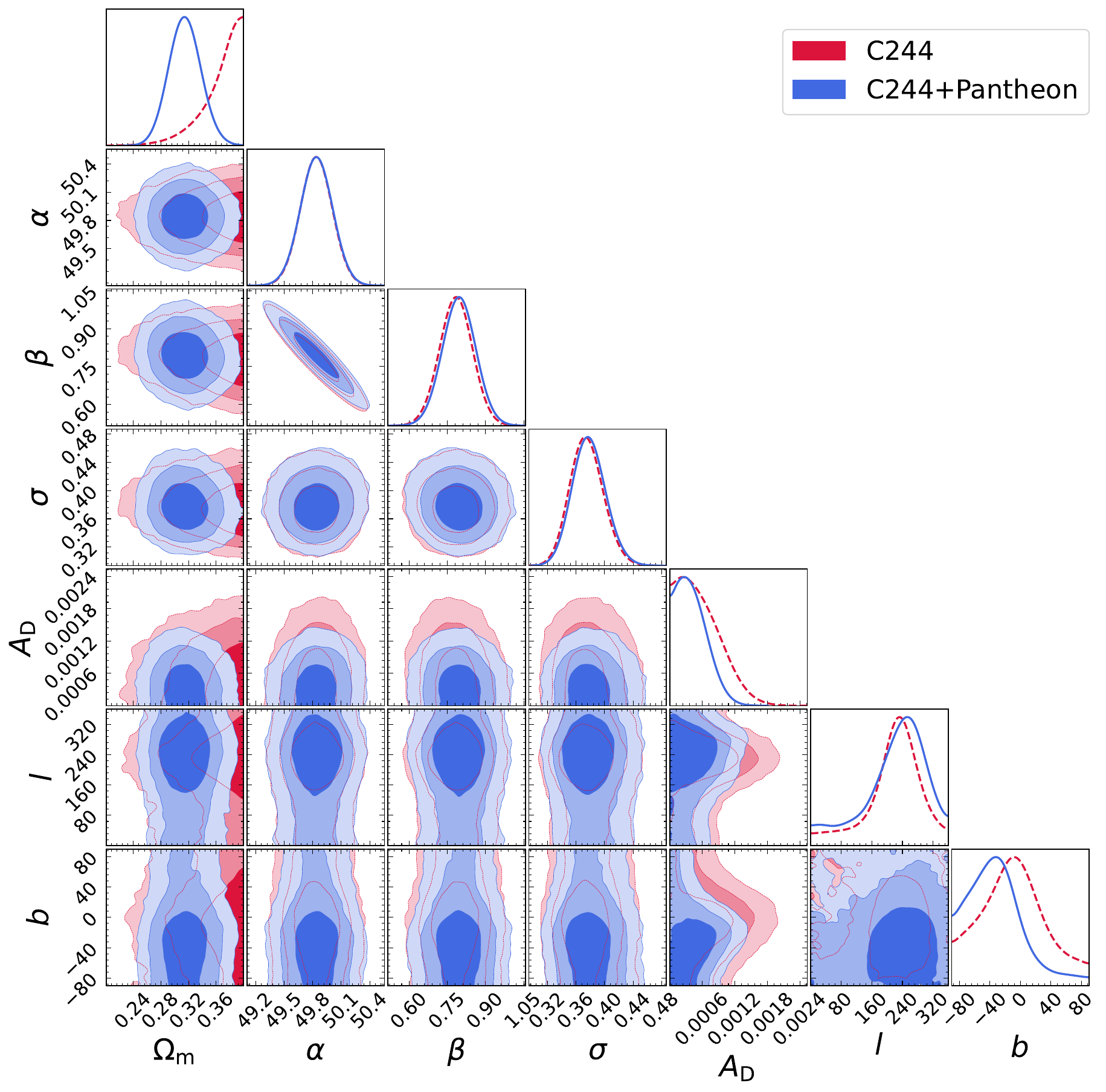}
		\caption{Marginalized posterior distributions of parameters in the dipole-modulated $\Lambda$CDM model. The red and blue curves represent the results derived from the C244 and C244+Pantheon samples, respectively.}
		\label{fig_L}
	\end{center}
\end{figure*}

\begin{figure*}
	\begin{center}
		\includegraphics[width=15cm]{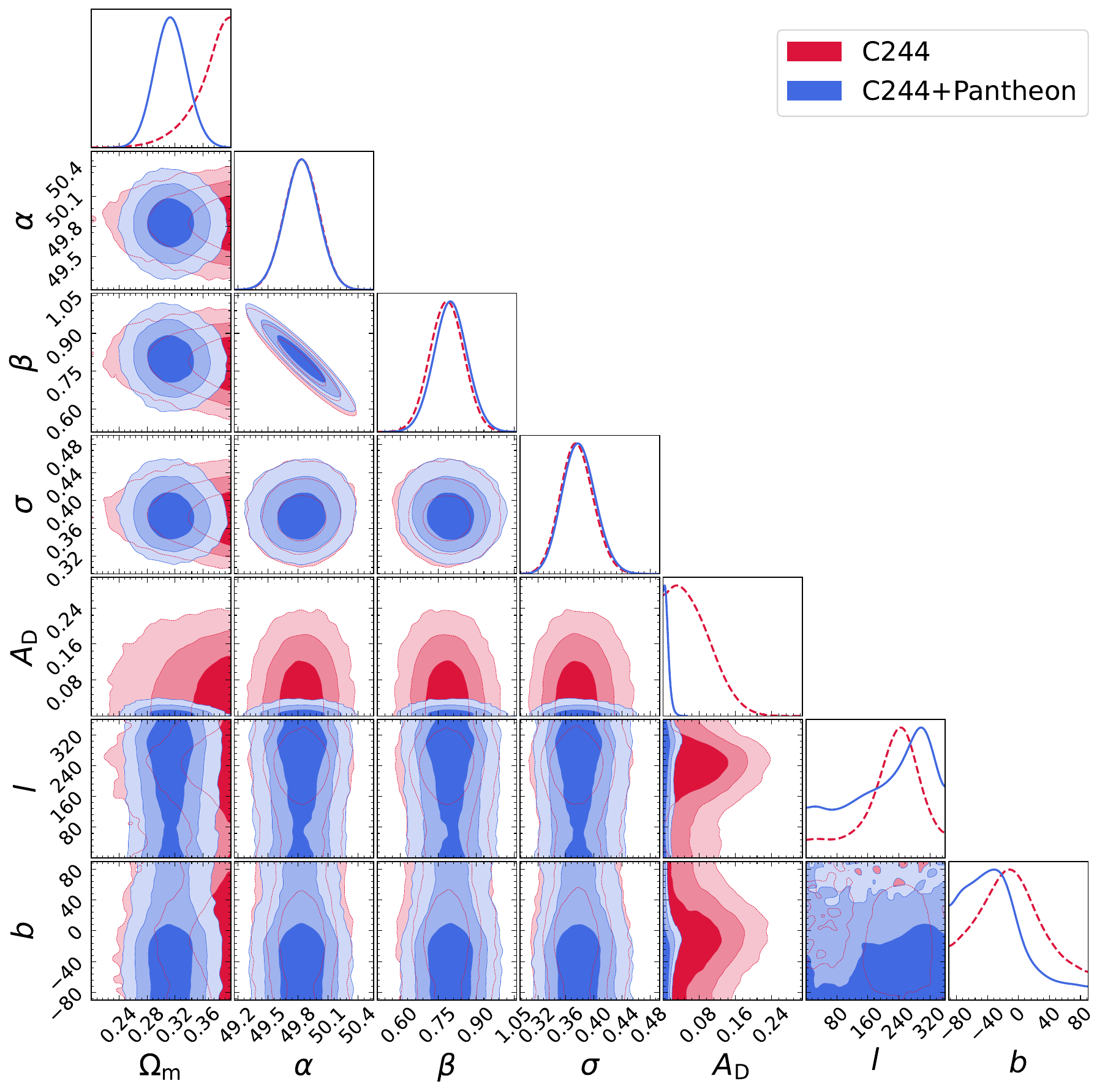}
		\caption{Marginalized posterior distributions of parameters in the Finslerian cosmological model. The red and blue curves represent the results derived from the C244 and C244+Pantheon samples, respectively.}
		\label{fig_F}
	\end{center}
\end{figure*}

The full parameter space is explored via the Markov Chain Monte Carlo (MCMC) method using the Python package \texttt{emcee}\footnote{https://emcee.readthedocs.io/en/stable/} \citep{Foreman-Mackey:2012any}. The flat priors for the free parameters in this work are set as follows: $\Omega_\mathrm{m}\in[0.2,0.4]$, $\alpha\in[48,51]$, $\beta\in[0.5,1.1]$, $\sigma\in[0.25,0.5]$, $A_\mathrm{D}\in[0,1]$, $l\in[0^\circ,360^\circ]$, $b\in[-90^\circ,90^\circ]$, $\mathcal{M}\in[23.5,24]$. 
We present the results in Figs. \ref{fig_L} and \ref{fig_F}, and summarize them in Table \ref{table_1}. To avoid redundancy, we do not show the results of the nuisance parameter $\mathcal{M}$ from the Pantheon sample.

\begin{table*}
	\large
	\begin{center}
		\caption{Best-fit parameters with 68\% confidence level (CL) are summarized for the dipole-modulated $\Lambda$CDM model and the Finslerian cosmological model. For parameters with only an upper or lower bound, the 95\% CL is reported instead.}
		\resizebox{\textwidth}{!}{
			\renewcommand{\arraystretch}{1.3}
			\setlength{\tabcolsep}{1.0mm}{
				\begin{tabular}{ccccccccc}
					\hline 
					Model & Data & $\Omega_\mathrm{m}$ & $\alpha$ & $\beta$ & $\sigma$ & $10^{3}A_\mathrm{D}$ & $l[^{\circ}]$ & $b[^{\circ}]$ \\
					\hline
					${\rm \Lambda}$CDM 
					& C244 & $ - $ & $ 49.854_{ -0.173}^{ +0.133}$ & $ 0.785_{ -0.058}^{ +0.060}$ & $ 0.373_{ -0.020}^{ +0.023}$ & $< 1.241$ & $ 232.49_{ -56.96}^{ +57.18}$ & $ -8.33_{ -47.86}^{ +36.73}$ \\
					& C244+Pantheon & $ 0.318_{ -0.025}^{ +0.018}$ & $ 49.840_{ -0.151}^{ +0.154}$ & $ 0.789_{ -0.053}^{ +0.066}$ & $ 0.380_{ -0.024}^{ +0.018}$ & $< 0.891$ & $ 255.01_{ -80.27}^{ +56.83}$ & $ -29.13_{ -45.30}^{ +21.53}$  \\
					& Pantheon & $ 0.296_{ -0.020}^{ +0.024}$ & $ - $ & $ - $ & $ - $ & $< 1.142$ & $ 309.10_{ -125.40}^{ +86.30}$ & $ -33.65_{ -56.02}^{ +15.28}$ \\
					& A118+Pantheon & $ 0.298_{ -0.022}^{ +0.022}$ & $ - $ & $ - $ & $ - $ & $ < 1.128$ & $ 276.36_{ -77.33}^{ +56.04}$ & $ -37.30_{ -40.93}^{ +20.49}$ \\
					\hline
					Finslerian 
					& C244 & $ - $ & $ 49.841_{ -0.158}^{ +0.154}$ & $ 0.794_{ -0.070}^{ +0.050}$ & $ 0.375_{ -0.022}^{ +0.021}$ & $< 145.741$ & $ 249.43_{ -75.27}^{ +55.27}$ & $ -12.20_{ -50.82}^{ +39.94}$  \\
					& C244+Pantheon & $ 0.314_{ -0.023}^{ +0.020}$ & $ 49.841_{ -0.159}^{ +0.150}$ & $ 0.802_{ -0.063}^{ +0.056}$ & $ 0.380_{ -0.024}^{ +0.018}$ & $< 18.916$ & $ 301.47_{ -140.83}^{ +56.55}$ & $ -26.43_{ -56.66}^{ +14.91}$  \\
					& Pantheon & $ 0.296_{ -0.022}^{ +0.022}$ & $ -$ & $ -$ & $ -$ & $ < 19.181 $ & $ 305.45_{ -148.38}^{ +57.44}$ & $ -35.46_{ -50.27}^{ +19.96}$ \\
					& A118+Pantheon & $ 0.297_{ -0.020}^{ +0.024}$ & $ -$ & $ -$ & $ -$ & $ < 20.251$ & $ 301.83_{ -133.36}^{ +61.21}$ & $ -33.64_{ -45.35}^{ +23.20}$ \\
					\hline
				\end{tabular}
			}
	    }\label{table_1}
	\end{center}
\end{table*}

In the dipole-modulated $\Lambda$CDM model, the Combo correlation constraints from the C244 sample are determined to be $\alpha=49.854^{+0.133}_{-0.173}$, $\beta=0.785_{-0.058}^{+0.060}$, and $\sigma=0.373_{-0.020}^{+0.023}$. The dipole amplitude $A_\mathrm{D}$ yields an upper limit of $A_\mathrm{D}<1.241\times 10^{-3}$ at the 95\% CL, implying that the dipole anisotropy is very weak. Additionally, the dipole direction points toward $(l,b)=(232.49^{{\circ}+57.18^{\circ}}_{~-56.96^{\circ}}, -8.33^{{\circ}+36.73^{\circ}}_{~-47.86^{\circ}})$. For the Finslerian cosmological model, the results for the parameters in the Combo correlation are almost identical to those obtained in the dipole-modulated $\Lambda$CDM model. A negligible dipole anisotropy is observed, with an upper limit of $A_\mathrm{D}<0.146$ at the 95\% CL. The dipole direction is $(l,b)=(249.43^{{\circ}+55.27^{\circ}}_{~-75.27^{\circ}}, -12.20^{{\circ}+39.94^{\circ}}_{~-50.82^{\circ}})$.

We combine the C244 sample with the Pantheon sample to obtain joint cosmological constraints. In the dipole-modulated $\Lambda$CDM model, the Combo correlation constraints from the joint dataset are determined to be $\alpha=49.840_{-0.151}^{+0.154}$, $\beta=0.789_{-0.053}^{+0.066}$, and $\sigma=0.380_{-0.024}^{+0.018}$. The matter density $\Omega_\mathrm{m}$ is $0.318_{-0.025}^{+0.018}$. The dipole amplitude $A_\mathrm{D}$ yields an upper limit of $A_\mathrm{D}<0.891\times 10^{-3}$ at the 95\% CL and the dipole direction points to $(l,b)=(255.01^{{\circ}+56.83^{\circ}}_{~-80.27^{\circ}}, -29.13^{{\circ}+21.53^{\circ}}_{~-45.30^{\circ}})$. For the Finslerian cosmological model, the results of the parameters in the Combo correlation are almost identical to those mentioned above. The matter density $\Omega_\mathrm{m}$ is $0.314_{ -0.023}^{ +0.020}$. The dipole amplitude $A_\mathrm{D}$ yields an upper limit of $A_\mathrm{D}<0.019$ at the 95\% CL, with the corresponding direction pointing to $(l,b)=(301.47^{{\circ}+56.55^{\circ}}_{~-140.83^{\circ}}, -26.43^{{\circ}+14.91^{\circ}}_{~-56.66^{\circ}})$.

\begin{figure*}[t]
	\centering
	\begin{subfigure}[b]{0.48\textwidth}
		\centering
		\includegraphics[width=\textwidth]{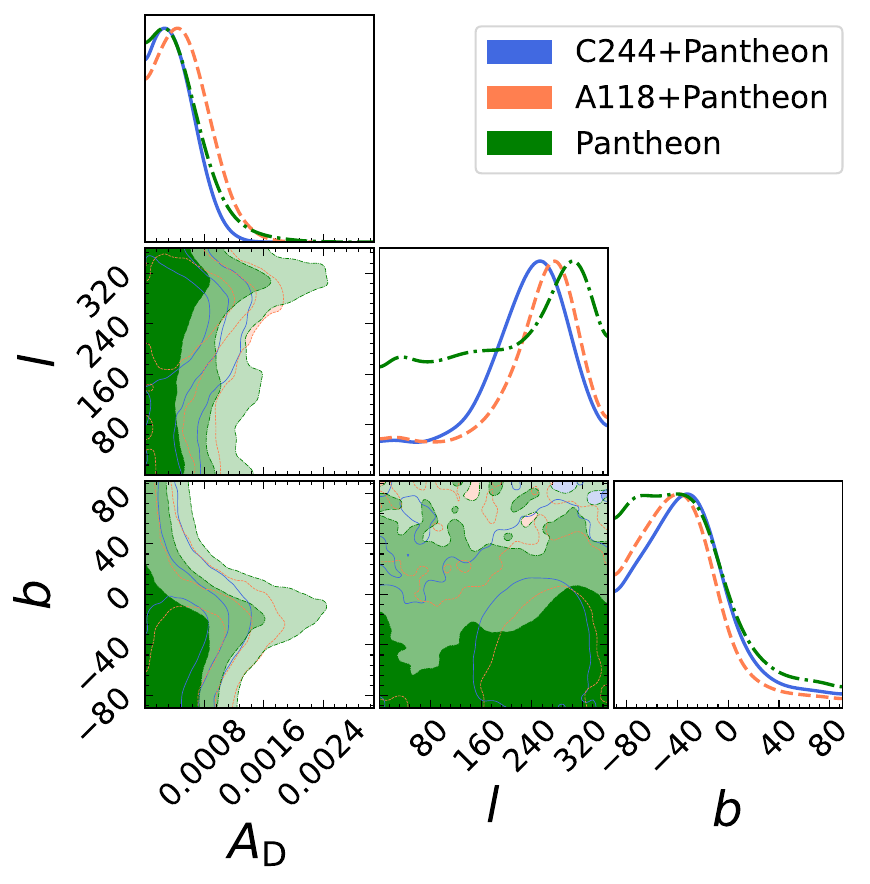}
		\caption{Anisotropic signals in the dipole-modulated $\Lambda$CDM model.}
		\label{fig_lb_A}
	\end{subfigure}
	\hfill
	\begin{subfigure}[b]{0.48\textwidth}
		\centering
		\includegraphics[width=\textwidth]{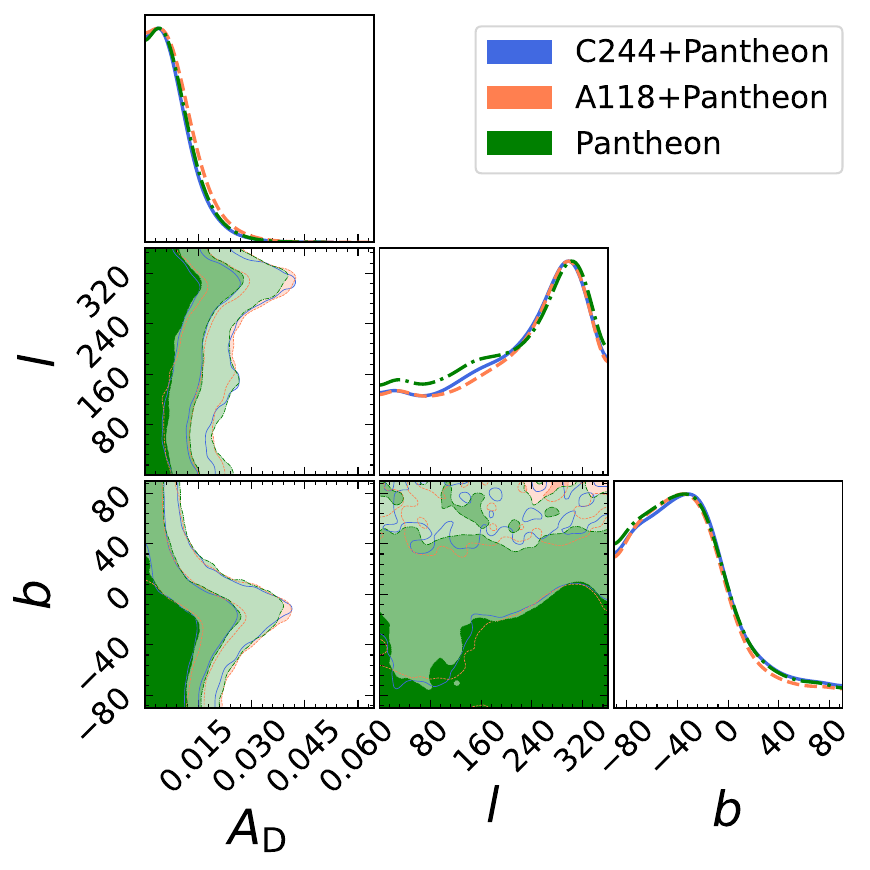}
		\caption{Anisotropic signals in the Finslerian cosmological model.}
		\label{fig_lb_B}
	\end{subfigure}
	\caption{Marginalized posterior distributions for the parameters of anisotropic signals. Panel (a) shows the results in the dipole-modulated $\Lambda$CDM model, and panel (b) shows the results in the Finslerian cosmological model. The blue and green lines denote the results obtained from the C244+Pantheon and Pantheon samples, respectively. The orange lines denote the results obtained from the A118+Pantheon sample, which are taken from our previous work \citep{Zhao:2022dwy}.}
	\label{fig_lb}
\end{figure*}

Figure \ref{fig_lb} displays the anisotropic signals derived from the C244+Pantheon, A118+Pantheon, and Pantheon samples. The results for the A118+Pantheon sample, which are also shown in Table \ref{table_1}, are taken from our previous work \citep{Zhao:2022dwy}. The results indicate that the dipole amplitudes for all datasets are small, suggesting that the anisotropic signals in these datasets are very weak. In the dipole-modulated $\Lambda$CDM model, when the Pantheon sample is combined with the two GRB samples, respectively, the results for the Galactic longitude $l$ show obvious changes. The uncertainty in the longitude $l$ is reduced by approximately $40\%$. The addition of the A118 sample induces a shift of $32.74^{\circ}$ in the longitude $l$. With approximately twice the sample size of the A118 sample, the C244 sample shifts the longitude $l$ by $54.09^{\circ}$. Additionally, under the Finslerian cosmological model, the dipole directions determined from these three datasets are highly consistent with each other.

\subsection{Hemisphere comparison}\label{R_HC}
\begin{figure*}
	\begin{center}
		\includegraphics[width=17cm]{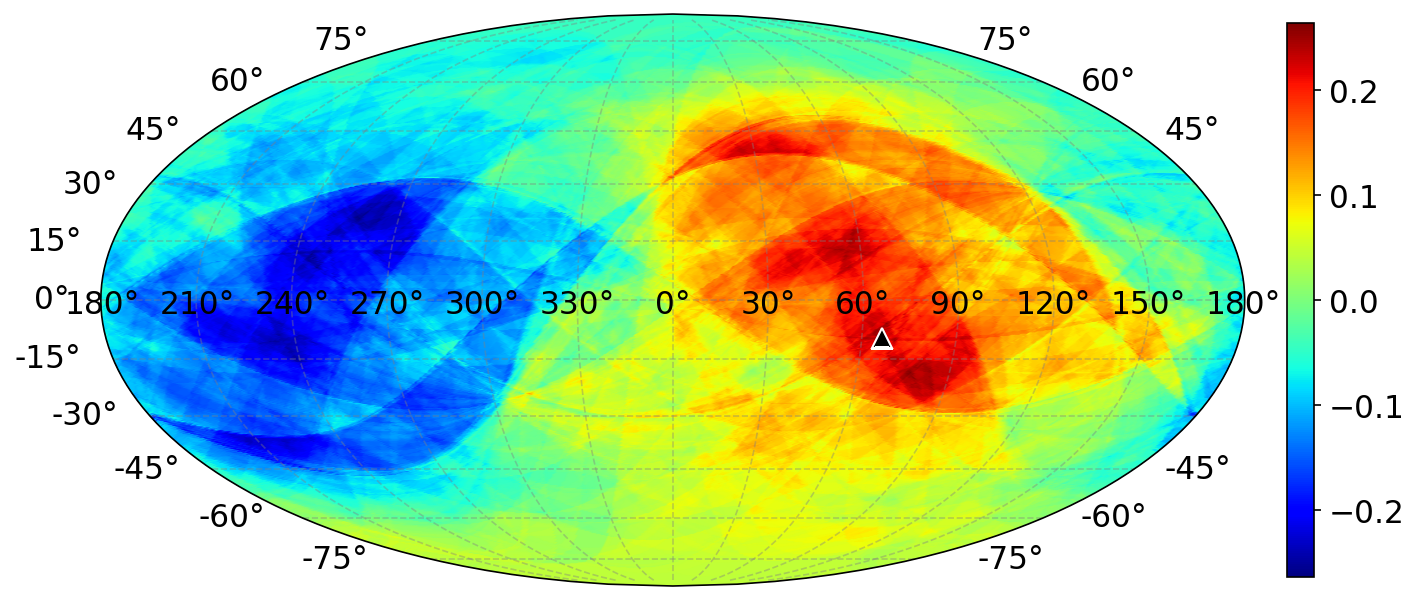}
		\caption{Pseudo-color map of the AL from the C244+Pantheon sample in Galactic coordinates. The triangle marks the position of $\mathrm{AL}_\mathrm{max}$.}
		\label{fig_AL}
	\end{center}
\end{figure*}

In the HC method, we use the Python package \texttt{healpy}\footnote{http://healpix.sourceforge.net} \citep{Gorski:2004by, Zonca:2019vzt} to generate random directions with $N_\mathrm{side}=128$. The total number of random directions is $12 \times N_\mathrm{side}^2=196,608$. In the analysis, we fix the Combo correlation parameters at $\alpha=49.80$, $\beta=0.791$, and $\sigma=0.375$. These values are adopted from the flat $\Lambda$CDM model. The C244 sample is combined with the Pantheon sample to obtain joint cosmological constraints. We minimize $\chi^2=\chi^2_\mathrm{GRB}+\chi^2_\mathrm{Pantheon}$ to determine the best-fitting $\Omega_\mathrm{m}$ and its $1\sigma$ uncertainty for the subsets corresponding to the ``up'' and ``down'' hemispheres. Figure \ref{fig_AL} shows the pseudo-color map of AL from the C244+Pantheon sample in Galactic coordinates. The $\mathrm{AL}_\mathrm{max}$ is $0.264 \pm 0.057$ and the corresponding direction is $(l,b)=(66.45^{{\circ}+31.29^{\circ}}_{~-51.68^{\circ}}, -9.59^{{\circ}+53.00^{\circ}}_{~-17.69^{\circ}})$. In the preferred direction, the best-fitting values of $\Omega_\mathrm{m}$ in the ``up'' and ``down'' hemispheres are $\Omega_\mathrm{m, \mathrm{u}}=0.346 \pm 0.023$ and $\Omega_\mathrm{m, \mathrm{d}}=0.266 \pm 0.026$, respectively.

Table \ref{table_2} summarizes the results obtained from the joint datasets (C244+Pantheon and A118+Pantheon) and the individual Pantheon sample. The two joint datasets produce highly consistent values for $\mathrm{AL}_\mathrm{max}$. When the Pantheon sample is combined with the two GRB samples, respectively, the values of $\mathrm{AL}_\mathrm{max}$ decrease significantly. Nevertheless, all values of $\mathrm{AL}_\mathrm{max}$ are consistent within the $1\sigma$ uncertainty.

\begin{table*}
	\large
	\begin{center}
		\caption{Results of the HC method obtained from the joint datasets (C244+Pantheon and A118+Pantheon) and the individual Pantheon sample.}
		\resizebox{\textwidth}{!}{
			\renewcommand{\arraystretch}{1.5}
			\setlength{\tabcolsep}{2.0mm}{
				\begin{tabular}{cccccc}
					\hline 
					Sample & $\Omega_{m, u}$ & $\Omega_\mathrm{m, d}$ & $\mathit{AL}_\mathrm{max}$ & $l[^{\circ}]$ & $b[^{\circ}]$ \\
					\hline
					C244+Pantheon & $0.346 \pm 0.023$ & $0.266 \pm 0.026$ & $0.264 \pm 0.057$ & $66.45_{ -51.68}^{ +31.29}$ & $ -9.59_{ -17.69}^{ +53.00}$ \\
					A118+Pantheon & $0.344 \pm 0.023$ & $0.266 \pm 0.028$ & $0.257 \pm 0.060$ & $82.97_{ -61.88}^{ +52.73}$ & $ -15.09_{ -13.54}^{ +60.09}$ \\
					Pantheon & $0.313 \pm 0.023$ & $0.217 \pm 0.029$ & $0.361 \pm 0.070$ & $123.05_{ -4.22}^{ +11.25}$ & $ 4.78_{ -8.36}^{ +1.80}$ \\
					\hline
				\end{tabular}
			}
	    }\label{table_2}
	\end{center}
\end{table*}

\begin{figure*}
	\begin{center}
		\includegraphics[width=17cm]{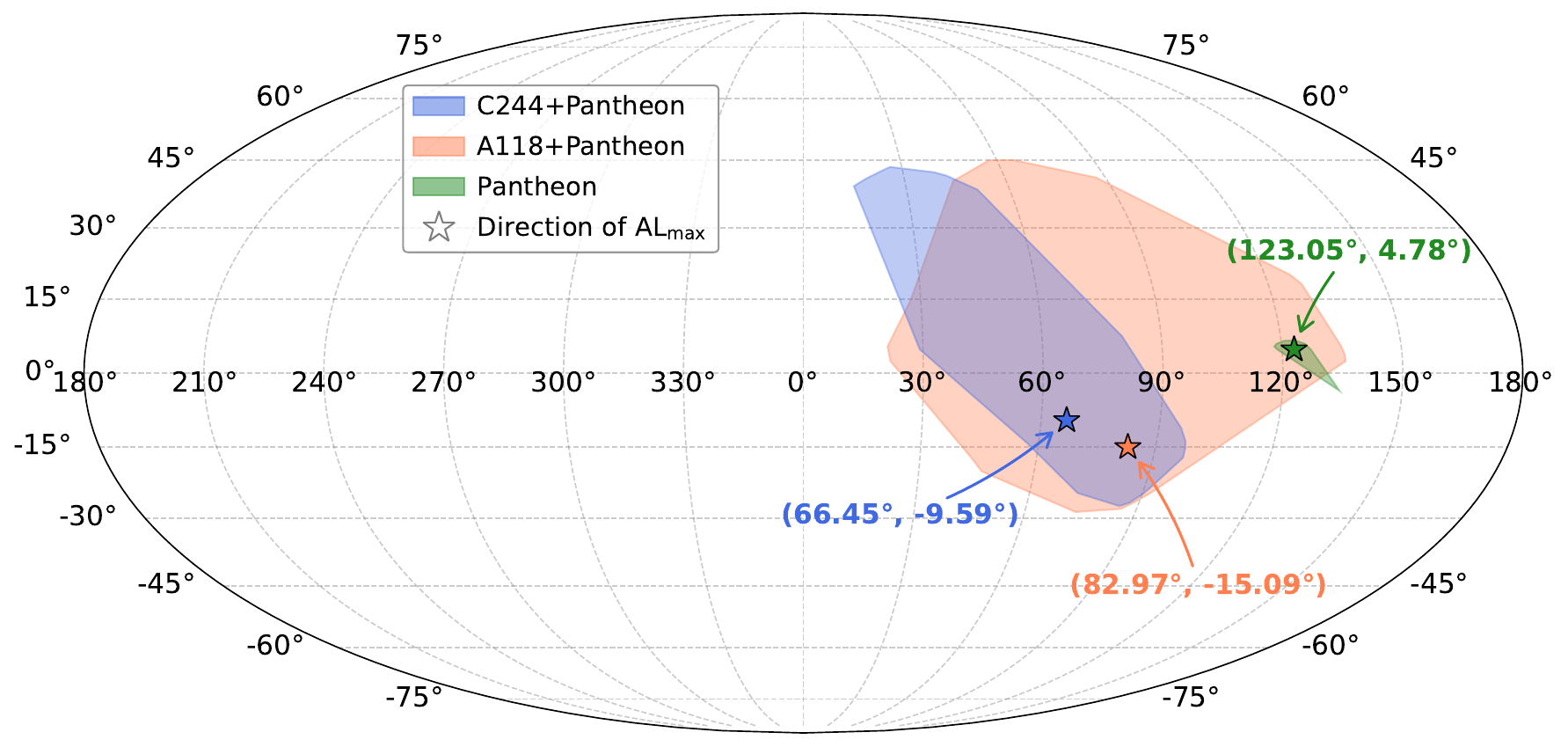}
		\caption{Preferred direction corresponding to $\mathrm{AL}_\mathrm{max}$ along with its $1\sigma$ uncertainty region. The blue, orange, and green shaded areas indicate the results from the C244+Pantheon, A118+Pantheon, and Pantheon samples, respectively. The star markers represent the preferred directions corresponding to $\mathrm{AL}_\mathrm{max}$ for each sample.}
		\label{fig_lb_1sigma}
	\end{center}
\end{figure*}

In Figure \ref{fig_lb_1sigma}, we plot the preferred direction corresponding to $\mathrm{AL}_\mathrm{max}$ along with its $1\sigma$ uncertainty region. The preferred direction derived from the A118+Pantheon sample is consistent with that from the Pantheon sample within $1\sigma$ uncertainty. Clearly, the preferred direction from the C244+Pantheon sample does not exhibit such consistency, deviating beyond the $1\sigma$ uncertainty. As the number of GRBs increases from 118 to 244, the preferred direction changes significantly.

\section{Conclusion}\label{Conclusion}
We employed the sample of 244 GRBs with the Combo correlation to test the cosmic anisotropy in both the dipole-modulated $\Lambda$CDM model and the Finslerian cosmological model. The anisotropic signal in the C244 sample is negligible. By combining the C244 sample with the Pantheon sample, we obtained joint cosmological constraints. 

For the DF method, we found that combining the C244 sample with the Pantheon sample in the dipole-modulated $\Lambda$CDM model reduces the uncertainty in longitude $l$ by approximately $40\%$. Furthermore, the incorporation of the C244 sample leads to the $54.09^\circ$ shift in the best-fitting longitude $l$. Compared with the A118+Pantheon sample, the shift in longitude $l$ increases by an additional $21.35^\circ$ as the number of GRBs increases from 118 to 244. In the Finslerian cosmological model, the dipole directions determined from the C244+Pantheon, A118+Pantheon, and Pantheon samples are highly consistent with each other.

In the HC method, for the C244+Pantheon sample, we found $\mathrm{AL}_\mathrm{max}=0.264 \pm 0.057$ and the corresponding direction is $(l,b)=(66.45^{{\circ}+31.29^{\circ}}_{~-51.68^{\circ}}, -9.59^{{\circ}+53.00^{\circ}}_{~-17.69^{\circ}})$. Compared to the Pantheon-only result, the value of $\mathrm{AL}_\mathrm{max}$ decreases significantly. Nevertheless, these two values of $\mathrm{AL}_\mathrm{max}$ remain consistent within $1\sigma$ uncertainty. However, we found that the preferred direction derived from the C244+Pantheon sample deviates from the Pantheon-only result by more than $1\sigma$. In contrast, the preferred direction from the A118+Pantheon sample is consistent with the Pantheon-only result within $1\sigma$ uncertainty. These results confirm that the preferred direction changes significantly as the number of GRBs increases from 118 to 244.

Our results show that GRBs can suppress the fake anisotropic signals induced by inhomogeneous spatial distributions. Therefore, we believe that GRBs have the potential to provide a reliable probe of cosmic anisotropy. Currently, GRBs still face several limitations as cosmological probes, such as limited sample sizes and a shortage of low-redshift sources. Space missions such as the currently successfully operating SVOM \citep{Bernardini:2021}, along with proposed future missions like THESEUS \citep{Amati:2018goh}, are expected to provide much larger and more precise GRB samples. These advanced observations will significantly improve the statistical precision of GRB-based cosmological constraints.

\backmatter
\bmhead{Acknowledgements}
J.-Q.X. is supported by the National Natural Science Foundation of China, under grant Nos. 12473004 and 12021003, the National Key R\&D Program of China, No. 2020YFC2201603.


\bibliography{sn-bibliography}% common bib file

\end{document}